\documentclass{comjnl}

\usepackage{amsmath}
\usepackage{algorithm}
\usepackage{algorithmic}  
\usepackage[algo2e,ruled]{algorithm2e} 
\usepackage{graphicx}
\usepackage{float}
\usepackage{enumerate}
\usepackage{array}
\usepackage{amssymb}
\usepackage{color}
\usepackage{soul}
\usepackage{amsmath}
\usepackage{caption,subcaption}
\usepackage{lipsum}

\usepackage{graphicx}
\usepackage{float,enumitem}
\usepackage{array}
\usepackage{amssymb}
\usepackage{color}
\usepackage{soul}
\usepackage{amsmath}
\usepackage{caption}
\usepackage{makecell}
\usepackage{multirow}      
\usepackage{multicol}

\begin{document}

\title[Calendar-based User Behavior Modeling]{CalBehav: A Machine Learning based Personalized Calendar Behavioral Model using Time-Series Smartphone Data}

%\title[Calendar-based User Behavior Modeling]{CalBehav: Calendar-based User Behavior Modeling using Time-Series Smartphone Data}

%\title[Modeling User Behavior]{Modeling User Behavior for Calendar Schedules using Time-Series Smartphone Data}
\author{Iqbal H. Sarker$^{1,2*}$, Alan Colman$^1$, Jun Han$^1$, A.S.M. Kayes$^3$ and Paul Watters$^3$}
\affiliation{$^1$ Swinburne University of Technology, \\ Melbourne, VIC-3122, Australia.\\
$^2$ Chittagong University of Engineering and Technology, \\ Chittagong-4349, Bangladesh.\\
$^3$ La Trobe University, \\ Melbourne, VIC-3086, Australia.\\
} \email{{\{*Corresponding author: msarker@swin.edu.au \}}}

\shortauthors{Sarker et al.}

\received{00 January 2009}
\revised{00 Month 2009}

%\category{C.2}{Computer Communication Networks}{Computer Networks}
%\category{C.4}{Performance of Systems}{Analytical Models}
%\category{G.3}{Stochastic Processes}{Queueing Systems}
%\terms{Internet Technologies, E-Commerce}
\keywords{user behavior modeling; machine learning; mobile data analytics; data science; calendar; smartphone; time-series; personalization; IoT and mobile services; intelligent systems; \\ (Preprint version of The Computer Journal, Oxford Unviersity Press, Uk.) }

%\footnotetext{This paper is an extension version of the DSAA'2016 special session paper ``Evidence-Based Behavioral Model for Calendar Schedules of Individual Mobile Phone Users''\cite{sarker2016evidence}}

\begin{abstract}
The electronic calendar is a valuable resource nowadays for managing our daily life appointments or schedules, also known as events, ranging from professional to highly personal. Researchers have studied various types of calendar events to predict smartphone user behavior for incoming mobile communications. However, these studies typically do not take into account \textit{behavioral variations} between individuals. In the real world, smartphone users  can differ widely from each other in how they respond to incoming communications during their scheduled events. Moreover, an individual user may respond the incoming communications differently in different contexts subject to what type of event is scheduled in her personal calendar. Thus, a \textit{static} calendar-based behavioral model for individual smartphone users does not necessarily reflect their behavior to the incoming communications. In this paper, we present a \textit{machine learning} based context-aware model that is \textit{personalized} and \textit{dynamically} identifies individual's dominant behavior for their scheduled events using logged \textit{time-series} smartphone data, and shortly name as \textit{``CalBehav''}. The experimental results based on real datasets from calendar and phone logs, show that this data-driven personalized model is more effective for intelligently managing the incoming mobile communications compared to existing calendar-based approaches.
\end{abstract}

\maketitle

\section{Introduction}
The electronic calendar has become the standard technology for personal event organization and management \cite{lovett2010calendar}. People typically use calendars, such as Google Calendar and Microsoft Outlook, to organize and manage various kinds of their daily activities and schedules from professional to highly personal, such as office related work (e.g., professional meetings), family events (e.g., picking kids from school), medical appointments, or public events etc. \cite{tungare2008exploratory}. For a particular event the calendar typically contains various attributes of the event, such as event name (e.g., meeting), days-of-week (e.g., Monday), time-of-the-day (e.g., 11 a.m.), duration of the event (e.g., 1 hour), event type (e.g., recurring), occurrence frequency (e.g., weekly) etc. Thus, the electronic calendar can be used as a potential source of information about users' various day-to-day situations in their life. However, such calendar events do not provide information on an individual's behavior related to incoming mobile communications, e.g., responding incoming phone calls, during the events. An individual's phone call responding activities may vary from user-to-user according to their own preferences. These behaviors are recorded in their phone log as a time-series. In this paper, we present ``CalBehav''which correlates personal calendar information with phone response behavior as recorded in the phone log to model an individual's behavioral patterns utilizing their smartphone log data.

Researchers have studied about calendar-based context-aware systems for smartphone users and corresponding contextual information in order to model and predict their behavior for mobile communication. A static event-behavior mapping based on the \textit{existence of an event} in the calendar or the \textit{name of an event} using special keywords and corresponding \textit{predefined rules} are used in these systems. For instance, if an event exists in the calendar of an individual, e.g., from 10 a.m. to 11 a.m. on Monday, it is highly assumed that the user is unavailable to answer the incoming calls during that time period \cite{khalil2005improving}. However, such calendar-based unavailability for responding their incoming phone calls may vary from user-to-user depending on event characteristics as well. For instance, one's behavior during the event `meeting' may not be similar with the event `lunch-break'. Unlike the event existence-based prediction model, researchers have also studied about the particular \textit{name of an event} based on a collection of special case words (e.g., meeting, lecture etc) in order to specify user behavior \cite{dekel2009minimizing}. However, such keyword or event name based systems are not individualized behavior-oriented. In the real world, the smartphone users are well differ from each other in how they respond to incoming communications during their personal scheduled events. Moreover, an individual user may respond the incoming communications differently in different contexts subject to what type of event is scheduled in her personal calendar. Thus, a \textit{static} calendar-based behavioral model and corresponding predefined \textit{rules} for individual smartphone users do not necessarily reflect individuals' behavior to the incoming communications. 

Let's consider an example of multiple calendar events and corresponding call response behavior of a mobile phone user, Alice. She is a postgraduate research student. On Monday between 9 a.m. and 10 a.m. she has a calendar event name `meeting'  with her supervisors at office and she typically rejects the incoming calls during the period of meeting. In the evening, between 3 p.m. and 5 p.m. she has another calendar event `meeting' with her friends who are also working in the same research lab, at tea room for group discussion about their individual research problems. However, she typically accepts the incoming calls during that time period, particularly incoming calls from her family members or close friends. Thus, the \textit{existence of an event} indicating the particular time period or the \textit{name of an event} based on special words are inadequate to infer individuals' behavior for their various calendar events. Moreover, additional contextual information such as social relationship between individuals (e.g., friend) or particular location (e.g., at office) might have an impact on individual's phone usage behavior. In many cases the calendar event entered by the user will not provide sufficient semantic information about the events to infer behavior \cite{lee2010exploiting}. For instance, the user may hide the actual event's name and uses `busy' instead, for privacy issue \cite{thayer2012love} or habitually enters inappropriate events' information (e.g., location name instead of event's name) in the calendar \cite{lee2010exploiting}. Thus, it's very difficult to predict individual's phone-response behavior from a static mapping to nominal calendar events  without additional behavioral evidence.

To effectively model and predict an individual's phone response behavior during their calendar events, a personalized data-driven dynamic \textit{event-behavior mapping} is needed. As such smartphone time-series data can be used as \textit{behavioral evidence} of actual phone call activities, to dynamically predict such behavior by synchronizing with the temporal context of the calendar. Smartphones automatically log individual's phone call activities with corresponding contextual information, particularly, call related information, e.g., call date, specific call time, call type, call duration, the corresponding contact number of the caller \cite{sarker2016phone}. We use such contextual raw data as the evidence for inferring the call response behavior (accept $|$ reject $|$ missed) of individuals for incoming communications for their various scheduled events in calendar. 

As well as variations of behavior between individuals in the real world, an individual's phone call response behavior may not always be consistent for any given set of contextual parameters. For instance, one individual user may prefer to reject calls for a particular calendar event, e.g., meeting at office, where in the past she has rejected calls more than, say, 85\% of the time for that event - that is, at a confidence preference threshold of 85\%. Another individual user, on the other hand, may only allow of 100\% (maximum) confidence preference according to her behavioral consistency in her real world activities. Thus, we aim to identify the \textit{dominant behavior} \cite{sarker2017individualized} for a particular association of contexts related to individuals' various calendar appointments, and to generate a set of event related \textit{behavioral rules} according to their confidence preferences. This preference for setting this confidence level may vary from user-to-user.

In the area of machine learning and data science, both association rule learning and classification rule learning are the two well-known techniques to extract rules from a given dataset. A rule $(Con \Rightarrow Behav)$ represents an IF-THEN logical statement \cite{sarker2018BehavMiner}; where the antecedent ($Con$) represents contextual information related to calendar events of an individual mobile phone user, and the consequent ($Behav$) represents her call response behavior for the corresponding events. In order to generate individuals' behavioral rules for their calendar events, we take into account association rule learning rather than using classification rule learning techniques. The reason is that classification rule learning technique allows a rigid decision making without considering users' preferences. Consequently the discovered classification rules may not have a high predictive accuracy \cite{sarker2019recencyminer}. In contrast, association rule learning is a well-defined, deterministic task that discovers a set of association rules having support and confidence greater than a given preference. Thus, it allows individuals' to configure their own preferences for creating rules according to their unique behavioral patterns for their various scheduled events in their personal calendars.

In this paper, we address the above discussed issues and propose a machine learning based personalized calendar behavioral model ``CalBehav'' that uses smartphone time-series data as behavioral evidence and dynamically identifies the dominant behavior of individual mobile users for their various scheduled events, which significantly extends our earlier work \cite{sarker2016evidence}. In our approach, we first collect data from multiple sources such as individual's calendar information and their phone log data, where calendar represents event related information and phone log contains the evidence of their behavioral patterns for the events. After that, we dynamically do a data-driven event-behavior mapping by synchronizing the temporal context utilizing the data of both calendar events and phone log records of individuals. Once the dynamic event-behavior mapping has been done, we then generate a concise set of behavioral association rules by taking into account the relevant contexts for their calendar events. As the behavior of different individuals are not identical in the real word, such discovered rules may differ from user-to-user or event-to-event according to their behavioral patterns.

The contributions of this paper are summarized as follows.

\begin{itemize}
	\item We design a data-driven event-behavior mapping based on temporal context utilizing individual's datasets collected from multiple data sources.
	
	\item We build a machine learning based personalized context-aware behavioral model for their calendar schedules and generate a concise set of behavioral association rules by taking into account their calendars' contexts.
	
	\item We have done experiments on individual's datasets and show that this technique is more effective while comparing with existing calendar-based approaches, in order to intelligently managing the incoming mobile communications.
\end{itemize}

The remainder of this paper is organized as follows. Section \ref{Related Work} provides a brief review of related work. In Section \ref{Problem-Statement}, we formally state our calendar-based behavior modeling problem for smartphone users. Section \ref{Key Features of Our Approach} highlights the key features of our context-aware calendar-based model. In Section \ref{Our Evidence-Based Model}, we present our proposed behavioral model ``CalBehav'' for calendar schedules of individual mobile users. We report the experimental results on real datasets in Section \ref{Experiments}. Some key observations of our study are summarized in Section \ref{Discussion}, and finally Section \ref{Conclusion and Future Work} concludes the paper and highlights future work.

\section{Related Work}
\label{Related Work}
A significant amount of research based on calendar information has been done for modeling user behavior for incoming mobile communication and corresponding context-aware systems. However, such research and systems do not consider personalized behavioral evidence based on actual phone call activities recorded in their phone log, i.e., in which contexts an individual user accepts, rejects or misses the incoming phone calls in various scheduled events in their personal calendar.

A number of authors use calendar information to determine user's unavailability for the purpose of predicting users' phone call behavior. An individual user is considered as unavailable if the user would not pick up the incoming call \cite{pielot2014large}. In the real world, such unavailability is not ignorable, as people are typically involved in several scheduled events in various day-to-day situations. Salovaara et al. \cite{salovaara2011phone} have conducted a study and show that 31\% of the incoming phone calls were unavailability related, i.e., the users are unavailable to answer the incoming phone calls for various reasons such as meetings, lectures, appointments, driving a vehicle, sleeping etc. Khalil et al. \cite{khalil2005improving} state that calendar events or entries are a good cue as to whether a person is available or unavailable for answering an incoming call. For instance, if a calendar has a meeting appointment from 1 p.m. to 2 p.m. for an individual user, it is highly assumed with a high degree of probability that the user is in a place with other persons and she is unavailable to answer the incoming call during that time period. The reason is that people typically want to avoid interruptions in their real world life as such interruptions may create embarrassing situation for her which not only create disturbance for her but also for the surrounding people for that meeting.

However, such calendar-based static unavailability solution for users may provide low prediction accuracy in some real world cases. Khalil et al. \cite{khalil2005improving} have conducted a survey of seventy-two phone users and found that the above calendar-based unavailability solution for incoming mobile communications gives low accuracy (62\%) for loosely structured home activities such as `lunch break', `watching TV', `homework', while gives high level of accuracy (93\%) for the structured events such as `lecture', `meeting', `appointment' etc. These special keywords or event names are not sufficient for our real life use cases of daily activities, need a richer set of keywords to capture the actual mobile usage behavior of the users \cite{dekel2009minimizing}. In many cases, the incoming phone call may not be disruptive even though the user is engaged in an ongoing task or in a social situation. The reason is that the incoming call is welcome for someone as it provides a mental break from the current task which is needed for her \cite{de2007should}. For instance, Rosenthal et al. \cite{rosenthal2011using} have shown through a survey of users that 35\% of the participants want to answer phone calls even though at work, while other participants do not want. Thus, a calendar event based static behavioral model is not applicable for personalized solution for predicting their behavior in mobile communications.

A number of authors have also designed context-aware call interruption management system that produces output whether an incoming call should be allowed to ring or not. For example, Dekel et al. \cite{dekel2009minimizing} have designed a context-aware system for minimizing mobile phone disruptions by taking into account user's unavailability utilizing their calendar information. In \cite{zulkernain2010mobile}, the authors also utilize the calendar information in order to design an intelligent context-aware interruption management system. Seo et al. \cite{seo2011pyp} have defined a set of policy rules based on users' scheduled events in their context-aware phone configuration management system in order to improve cell phone awareness for the end mobile users. In these approaches, the interruption handling rules follow a static behavioral model based on user's unavailability utilizing calendar information, i.e., the user is unable to answer the incoming calls whenever she is in an appointment scheduled in the calendar. In contrast, individual's phone call activities recorded in their phone log is a good resource for their behavioral evidence and to infer their personalized behavior dynamically \cite{sarker2016phone}.

Besides calendar events, a number of authors have considered additional contextual information in order to infer user's behavior for mobile communication. Additional contexts can play a significant role to model user behavior as the calendar alone is limited as a `sensor' highlighted in Lovett et al. \cite{lovett2010calendar}. A probabilistic interruption management system based on multiple contexts has been proposed by Vilwock et al. \cite{vilwock2013system}. In their approach, they take into account calendar information, caller relationship etc. as contexts. Stern et al. \cite{stern2011preliminary} describe an approach of automatically detecting user's interruptibility, based on contexts such as location and calendar information. Sykes et al. \cite{sykes2014cloud} present a cloud-based interruption management system using multi-dimensional contexts where calendar schedules are used as environmental context. They mainly focus on the advantages of using cloud by taking into account the limitations of mobile devices' resources. However, such contexts used in different approaches may not provide the evidence of individual's phone call behavior in their different day-to-day situations. Among the data-driven approaches, recently, Sarker et al. \cite{sarker2019machine}, propose a machine learning based robust model for modeling mobile phone user behavior based on multi-dimensional contexts. In another study, Sarker at al. \cite{sarker2019classifications} have analyzed the effectiveness of various machine learning based models by taking into account temporal, spatial or social contexts for building relevant context-aware systems utilizing individual's smartphone data. However, in these data-driven approaches, individual's calendar schedules are not taken into account, in which we are interested in.

Unlike these works, in this paper, we present ``CalBehav'', i.e., a data-driven calendar-based personalized behavior model that dynamically maps individuals' calendar events using their time-series smartphone data, in order to identify their dominant behavior for their various scheduled events and to generate a set of corresponding behavioral association rules using machine learning technique. 

\section{Definitions and Problem Statement}
\label{Problem-Statement}

This section introduces main notions concerning calendar-based user behavior modeling using their time-series smartphone data and formally states personalized behavior modeling problem for their calendar events. In the following, the notions related to calendar-based user behavior modeling are defined and discussed with related examples. \\

\textbf{Definition 3.1 (Calendar Event).} \textit{Let, $Cal$ be an electronic calendar, e.g., Google Calendar, representing a number of entries of an individual user. A particular entry of the calendar $Cal$ is called a scheduled event for that user, if it contains the basic elements, such as event name, temporal information, event type etc. of a calendar appointment, i.e., attribute-value pair, called a feature or contextual information.} For instance if ``Meeting'' is a particular entry of $Cal$, then `event name' =``Meeting'' is represented as an attribute-value pair of the calendar. \\

\textbf{Definition 3.2 (Time-Series Smartphone Data).} \textit{Let, $T_{series}$ be a data feature and $Q$ its corresponding domain. $T_{series}$ is a sequence of data points ordered in time such that $T_{series} = (t_1, t_2,..., t_m)$, where $t_1, t_2,...,t_m$ are individual observations, each of which contains real-value data and $m$ is the number of observations in a time-series. A time-series mobile phone dataset $DS_{T}$ is a collection of records \cite{sarker2017individualized}, where}
\textit{
	\begin{enumerate}[label=(\roman*)]
		\item each record $r$ is a set of pairs $(t_i, value_i)$, where $t_i \in T_{series}$ that represents the timestamps, and $value \in Q$. For example, `2016-09-10 19:38:20' is a value of $t_i$, which represents the timestamps information in the format YYYY-MM-DD hh:mm:ss.
		\item each $t_i \in T_{series}$, also called attribute (temporal context), may occur at most once in any record, and
		\item each record has a particular user activity with mobile phones (e.g., reject phone call).
	\end{enumerate}
}

In the real world, the common incoming phone call activities of an individual mobile phone user are - (i) answering incoming phone call, i.e., `accept', (ii) decline the incoming phone call by the user, i.e., `reject', and (iii) the phone rings but the user misses the call, i.e., missed \cite{sarker2018Unavailability}. For each record, all the call related meta-data including temporal information are recorded. In order to produce the behavioral association rules of individual mobile phone users, we use the concept of \textit{dominant behavior} for a particular calendar event. Dominant behavior represents the ``maximum number of occurrences'' of a particular activity among a list of relevant activities for a given context \cite{sarker2017individualized}. In the following, the dominant behavior for calendar events is formally stated. \\

\textbf{Definition 3.3 (Dominant Behavior).} \textit{Let, $E$ is a particular calendar event and $ACT = \{act_1, act_2, ..., act_n\}$ be a list of diverse phone call activities of an individual mobile phone user for that event $E$, then $D = MAX({act_1, act_2, ..., act_n})$ is the dominant behavior of that calendar event.} For instance, in a particular calendar event, (say, seminar), a user has 85\% reject, 10\% accept, and 5\% missed call records, then `reject' will be the dominant behavior for that event. In another event (say, lunch), a user has 60\% accept, 40\% reject call records, then `accept' will be the dominant behavior for that event. \\

\textbf{Definition 3.4 (Behavioral Association Rule).} \textit{Let, $Con$ represents multi-dimensional contextual information related to calendar events of an individual user and $Behav$ is the corresponding behavior of that user. An association rule based on multi-dimensional contexts related to an individual's calendar events is represented in the form [$Con \Rightarrow Behav$], where $Con$ and $Behav$ are known as the antecedent and the consequent of the rule respectively.} An example of a behavioral association rule of an individual mobile phone user based on relevant multi-dimensional contexts is represented as: \\

[$calendar \; event \rightarrow meeting, social \; relationship \rightarrow friend  \Rightarrow phone \; call \; behavior \rightarrow reject$], where -
\textit{
	\begin{enumerate}[label=(\roman*)]
		\item Antecedent (Con): represents the multi-dimensional contextual information (e.g., $calendar \; event \rightarrow meeting, social \; relationship \rightarrow friend$) for the above example.
		\item Consequent (Behav): represents the corresponding phone call response behavior (e.g., $behavior \rightarrow reject$) of an individual mobile phone user for that contexts exist in the antecedent.
	\end{enumerate}
}

A behavioral association rule of an individual user must need to satisfy the support and confidence threshold preferred by the user. The confidence is calculated by the support count, which measures the accuracy of a rule. Higher confidence value of a rule ensures higher accuracy and vice-versa. A behavioral association rule is created if and only if it satisfies this confidence preference. For example, if the preferred confidence threshold of an individual user is (say, 80\%), then all the association rules that satisfy this threshold (e.g., $\ge$ 80\%) are discovered for that individual user. In the following, the definitions of support and confidence of an association rule is formally stated. \\

\textbf{Definition 3.5 (Support).}  \textit{Support is an indication of the frequency of occurrence, i.e., how many times the specific type of call has occurred in a dataset for a particular condition. Mathematically, a rule is an implication expression of the form  $A \Rightarrow C$, where A and C are disjoint data, i.e., $A \cap C=\phi$. If $Sup$ represents the support count and N is the total number of instances in dataset, then the formal definition of support is ${Sup (A \cup C)} / {N}$.} Rules that have a support greater than user-specified support is said to satisfy minimum support. A high support value means that the rule involves a great part of dataset. \\

\textbf{Definition 3.6 (Confidence).}  \textit{Let $A \Rightarrow C$ be a rule. Its confidence is given by $Sup (A \cup C) / Sup (A)$, where $Sup$ represents the support count mentioned above.} The confidence of a rule $A \Rightarrow C$ is the conditional probability of the user's behavior $C$ given the particular context $A$. It is measured as a percentage, e.g., 80\%. The generated behavioral association rules for various calendar events of an individual user must be \textit{non-redundant}, as the redundant generation unnecessarily increases the rule-size and consequently makes the resultant context-aware prediction model complex. In the following, we formally state the non-redundancy in behavioral association rules for individual's calendar events. \\

\textbf{Definition 3.7 (Non-Redundancy).} \textit{Let, two behavioral association rules for relevant calendar events are $R_1: A_1 \Rightarrow C_1$ and $R_2: A_2 \Rightarrow C_2$, we call the latter one redundant with the former one if $A_1 \subseteq A_2$ and $C_1 = C_2$. From this definition of redundancy, if we have a rule $R_g: A_1 \Rightarrow C_1$ and there is no other rule with additional context $A_1B_1 \Rightarrow C_2$ in existence such that confidence of $A_1B_1 \Rightarrow C_2$ is equal or larger than the confidence of $R_g: A_1 \Rightarrow C_1$ and $A_1 \subseteq A_1B_1$ and $C_1 = C_2$, then according to \cite{sarker2018mining} the rule $A_1B_1 \Rightarrow C_2$ is said to be non-redundant.} \\

\textbf {Problem Statement.} With the above definitions, the main problem we are addressing in this paper is stated as follows: \\

\textit{Given, a set of calendar events of an individual smartphone user and phone log data including time-series information with corresponding mobile phone activities of that user. Our goal is to design a dynamic event-behavior mapping based on that time-series data and to generate a set of behavioral association rules based on multi-dimensional contexts related to calendar events, which are non-redundant. Such rules are discovered at a particular level of confidence preferred by an individual, in order to effectively modeling individual mobile phone user behavior for their various events scheduled in the calendar. In this paper, we present ``CalBehav'' for solving this calendar-based user behavior problem.}

\section{CalBehav: Key Features}
\label{Key Features of Our Approach}
This section presents the key features of our data-driven approach to discover behavioral association rules of individual mobile phone users for their various scheduled events in the calendar. Our key features constitute the foundation of data-driven evidence-based behavioral model of individuals for their personal calendar events. The features are as follows:

\subsection{Behavioral Evidence}
Behavioral evidence for personalized calendar events is the most important feature of our CalBehav model. In the existing calendar-based systems, discussed in Section \ref{Related Work}, user behavior for mobile communication is predicted using calendar information based on a static event-behavior mapping. However, only the calendar information is not sufficient to effectively model and predict individual's actual call response behavior in their various day-to-day situations. The reason is that the calendar does not provide personalized behavioral evidence of the users. In the real world, each individual user may respond differently subject to what type of event is scheduled in her calendar. For instance, an individual's call response behavior during a `professional meeting' may be well different to response during a scheduled `tea-party' event. 

\begin{figure}
	\centering
	\includegraphics[width=\linewidth]{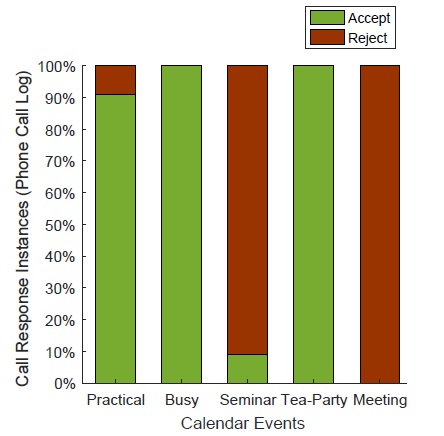}
	\caption{Behavioral phone call responses for various types of scheduled events in the calendar of a sample user}
	\label{fig:behavioral-evidence}
\end{figure}

Figure \ref{fig:behavioral-evidence} shows an example of the behavioral responses in phone call log (e.g., behavioral evidence) for various types of calendar events of a sample user. According to Figure \ref{fig:behavioral-evidence}, the user rejects most of the incoming calls ($>$90\%) during the events `Seminar' and `Meeting', while accepts most of the calls ($>$90\%) during the events `Practical' (represents lab work),`Busy' and `Tea-party'. As the phone call response behavior of individuals are not identical in the real world, such behavior may differ from user-to-user or event-to-event according to their preferences. Thus it is not appropriate to assume such diverse behaviors for various scheduled events of individuals. Therefore, in order to minimize the assumptions about phone call behavior, we incorporate a data-driven evidence-based behavioral model for various calendar events utilizing individual's phone log data, where phone data is used as the evidence of actual behavior of the corresponding user. 

In our CalBehav model, calendar is used as the indicator of user's activities over time, and phone call log to establish if there is a consistent mapping between types of calender event and actual phone call behavior. For instance, a user has two calendar events $E_1$ (lab work) from 10 a.m. to 11 a.m. and $E_2$ (lecture) from 2 p.m. to 3 p.m. on Mondays. In her phone log, most of the call instances are accepted during $E_1$ and rejected during $E_2$ on Mondays. According to the evidence of her past behavior for these two events, we conclude that her behavior is accept for event $E_1$ and reject for event $E_2$. Thus, the concept of data-driven evidence-based behavior mapping reflects the actual behavior of individuals for their various scheduled events in the calendar. 

\subsection{Generalization}
As we do not have any prior knowledge about individuals' phone call activities and corresponding recorded log data, it's difficult to assume for generating their behavioral rules for all scheduled events separately. The reason is that for some events, there may not have sufficient data evidence in phone log to produce rules for the corresponding events. For instance, a user has 10 nonrecurring events $E_1, E_2,...,E_{10}$ (presumably these are independent from each other) in her calendar. In her phone log, most of the incoming calls are rejected during the time period of $E_1, E_2,...,E_8$ and no evidence (data) is found during $E_9$ and $E_{10}$ as no calling activity occurs during these two events ($E_9$ and $E_{10}$). Since each of these nonrecurring events has its own contextual information, rules based on evidence may not be significant for all events, particularly for those events which has low evidence or no evidence at all (e.g., $E_9$ and $E_{10}$). In such cases, if we restrict producing rules for each event separately, meaningful rules may not be found for low evidence or evidence-less events. Thus, the concept of generalization helps to produce rules by taking into account `event type $\Rightarrow$ nonrecurring' as rule antecedent rather than individual events. In such cases, the generalized rule would be produced by taking into account the aggregated behavioral evidence (call response instances) of all nonrecurring events ($E_1, E_2,...,E_{10}$), e.g., if the event is nonrecurring then the user rejects the phone call that would be effective for all events in this example. As we do not produce any rules without behavioral evidence found in one's phone log, generalization plays a significant role to produce behavioral rules of individuals in our CalBehav model.

\subsection{Personalization}
Personalization is another significant feature of our CalBehav model. In the real world, people differ from each other in how they respond to incoming calls during their scheduled events, either unlike events (e.g., `meeting', `lunch') or like events (e.g., `meeting\_1', `meeting\_2') in individual's calendar. One user may not want to be interrupted during work but another user may want to answer. Say, a user X, was writing a paper when she received a call and appreciated the call as it forced her to take a needed mental break. One individual may be happy to take calls during an event – another may not. According to \cite{stern2011preliminary}, 24\% of cell phone users feel the need to answer a phone call when they are in a scheduled event, e.g., meeting and the rest of the users do not want to answer. For instance, in a `teacher-student meeting', the teacher accepts most of the incoming calls during the meeting while the student always rejects during that period. As phone call response behavior in mobile communication varies between individuals, personalization is a key feature in modeling mobile users behavior. However, the existing calendar based static behavioral model does not provide the automated personalized services learning individual user's preferences. As we infer the call response behavior for individual's calendar events utilizing their phone log data, our CalBehav model produces behavioral rules according to their unique behavioral patterns that are highly personalized.

In summary, by taking into account these features in our approach, we are able to resolve the issues of existing calendar-based approaches by producing individual's behavioral rules in order to make the calendar-based model more effective.

\section{Methodology: CalBehav}
\label{Our Evidence-Based Model}
In this section, we present our data-driven CalBehav model to extract a concise set of behavioral association rules of individual smartphone users for their various daily life events scheduled in their personal calendar.

\subsection{Approach Overview}
\begin{figure*}[htbp]
	\centerline{\includegraphics[width=.5\linewidth, height=9cm]{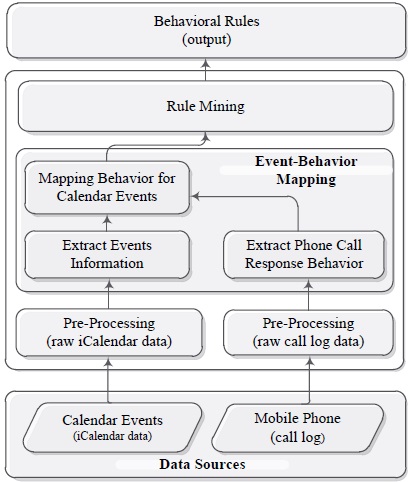}}
	\caption{An overview of our personalized calendar behavioral model ``CalBehav'' by employing rule-based machine learning technique utilizing multiple data sources.}
	\label{fig:overview}
\end{figure*}

The data-driven calendar-based personalized behavioral model ``CalBehav'' takes both the calendar and phone log data of an individual user as input data sources. Individuals' calendar information is used to determine one's scheduled events and corresponding call response behavior is discovered from their past call history recored in their phone log. Thus, from these two data sources, our approach is able to produce a set of behavioral association rules for the individual users for their various scheduled events by going through several processing steps. First, in pre-processing step, we extract the relevant contextual information for each event from the calendar data of an individual and corresponding call activity records for that individual from her phone log data. We then design a dynamic event behavior mapping by synchronizing the temporal context utilizing both the data sources and generate a personalized event-behavior list by taking into account all the relevant contexts, including temporal, spatial, social, and event related information. After that, we apply a rule-based machine learning technique, particularly, build an association generation tree for generating a set of concise behavioral association rules in order to achieve our goal. As users' preferences may vary from user-to-user in the real world, the set of rules that satisfy the user preferred confidence threshold is discovered for an individual user. Figure \ref{fig:overview} shows an overview of our machine learning based personalized calendar behavioral model CalBehav that discovers individualized behavioral association rules for their various daily life events scheduled in their personal calendar.

% Figure

\subsection{Data Pre-processing}
In this step, we pre-process Google provided iCalendar data source that contains the scheduled events' information of an individual. For each event, we extract the raw contextual data, e.g., event date, specific time, event duration, event name, event type, event occurrence frequency etc. that can be used to characterize an event of the calendar. Table \ref{Sample-iCal-data} shows a snippet of iCalendar data for a single event, where the event name is `Meeting' that occurs every two weeks on Thursday between 8 a.m. and 9 a.m.

\begin{table}
	\centering
	\caption{An example of iCalendar data for a sample scheduled event ``Meeting'' with relevant contextual information.}
	\label{Sample-iCal-data}
	\begin{tabular}{ |c| } 
		\hline
		BEGIN:VEVENT \\
		DTSTART;TZID=Australia/Sydney:20160602T080000 \\
		DTEND;TZID=Australia/Sydney:20160602T090000 \\
		RRULE:FREQ=WEEKLY;INTERVAL=2;BYDAY=TH \\
		DTSTAMP:20160531T090014Z \\
		UID:b75f7j27r30h8ijdql2tepab3s@google.com \\
		CREATED:20160531T085905Z \\
		DESCRIPTION: \\
		LAST-MODIFIED:20160531T085940Z \\
		LOCATION: \\
		SEQUENCE:1 \\
		STATUS:CONFIRMED \\
		SUMMARY:Meeting \\
		TRANSP:OPAQUE \\
		END:VEVENT \\
		\hline
	\end{tabular}
\end{table}

As we use multiple data sources in our model, we also pre-process individual's phone log data to extract and differentiate user's diverse incoming call responses into three categories, i.e., accept, reject and missed, which are the user behaviors we are interested in our CalBehav model. As both the accepting and rejecting calls are stored as incoming call type in the phone log, we distinguish accept and reject calls by using another parameter `call duration'. If the call type = `incoming' and call duration $=$ 0 (second), then the call has been rejected; If the call type = `incoming' and call duration $>$ 0 (second), then the call has been accepted. Missed call is another separated call type that is recored in phone log with call duration zero second. We also take into account data-centric social relational context generated from individual's unique contact number in phone log \cite{sarker2018DataCentricSocialContext}. For instance, mother's contact number represents one relationship while boss's contact number represents another relationship and so on.

\subsection{Dynamic Event-Behavior Mapping}
Syntactically, a phone call response behavioral model for a calendar event of individuals is a pair (E, B), where E is the event information in calendar and B is the phone call response behavior for that event. As we use E and B from different data sources in our approach, we do mapping event-behavior of individuals in order to modeling such call response behavior during an event E. To achieve our goal, we use the following related information of a calendar event and call response behavior of an individual user.

$Event \{E_{date}, T_{start}, T_{end}, E_{name}, E_{type}, E_{freq}\}$ 
where, $E_{date}$ represents event date, $T_{start}$ is the start time of an event, $T_{end}$ is the end time of that event, $E_{name}$ represents the event name, $E_{type}$ represents event type and $E_{freq}$ represents the event occurrence frequency.

$UserBehavior \{Call_{date}, Call_{T}, Call_{N}, Call_{B}\}$ 
where, $Call_{date}$ represents call response date, $Call_{T}$ is the specific time of the calling activity, $Call_{N}$ represents individual's unique contact number and $Call_{B}$ represents the phone call response behavior of an individual user.

As both data sources contain temporal information (date, time), we map the calendar events with individual's phone call response behavior based on this temporal information. To do this, we first synchronize the temporal information of both sources, e.g., taking into account the call response instances that are occurred during the time period of an event E. After that we do mapping event-behavior for all the phone call response instances during that event. For instance, say a calendar event `meeting' that occurs every Monday between 10 a.m. to 12 p.m. In order to map user behavior for this event, we take into account all the call responses from the phone log during the same time period on Monday.

% Figure
\begin{figure*}[htbp]
	\centerline{\includegraphics[width=.7\linewidth, height=8cm]{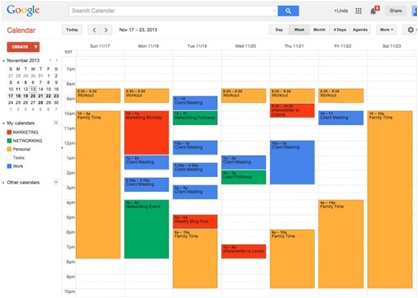}}
	\caption{An example of various daily life events scheduled in the calendar of a sample user.}
	\label{fig:calendar-events}
\end{figure*}

\begin{algorithm}
	\caption{Event-Behavior Mapping}
	\label{alg:calendar-behavior}
	\DontPrintSemicolon
	\SetAlgoLined
	\SetKwInOut{Data}{Data}
	\Data{event list: $E_{list}$, call list: $C_{list}$}
	\KwResult{event-behavior list: $EB_{list}$}
	
	\ForEach{event E in $E_{list}$}
	{  
		//extract event's temporal information \; \\
		$T_{event} \leftarrow$ extractTemporalInformation (E, $E_{list}$) \; \\
		//extract call records from $C_{list}$ during $T_{event}$ \; \\
		Bs $\leftarrow$ extractCallResponseBehavior ($T_{event}$, $C_{list}$) \;
		\ForEach{call response behavior B in Bs}
		{
			//event-behavior mapping with contexts\; \\
			EB $\leftarrow$ mapping (E, B) \; \\
			store EB in $EB_{list}$\;
		}	
	}	
	return $EB_{list}$\;
\end{algorithm}

The overall process for this is set out in Algorithm \ref{alg:calendar-behavior}. Input data include event list $E_{list}$ and phone call list $C_{list}$ and output data is the list of events and corresponding behavior. First, we extract the temporal information $T_{event}$ for each event E from $E_{list}$. We then extract all the call response instances during the time period of that event by synchronizing the temporal information in $C_{list}$. After that for each call response instances during $T_{event}$, we map the event information to its corresponding call response behavior. Finally, event-behavior list is generated by storing all the mapped data that is used for discovering individual's behavioral rules.

\subsection{Machine Learning based Rule Discovery}
In order to achieve our goal, we generate a concise set of behavioral association rules according to a particular level of confidence threshold preferred by individuals. Among the techniques of mining association rules, Apriori \cite{agrawal1994fast} is the most popular technique used in several application domains, particularly, market basket analysis. In order to generate user behavioral rules according to our goal, it produces a huge number of redundant associations and corresponding rules that make the behavior modeling approach complex. Thus, we employ the association generation tree based rule learning technique \cite{sarker2018mining} for mining the behavioral association rules of an individual user for their various calendar events. The reason for using this algorithm is that it produces a concise set of \textit{non-redundant behavioral association rules} for a particular confidence preference according to the precedence of contexts. Hence we briefly discuss our rule generating procedure step-by-step.

\subsubsection{Identifying the Precedence of Contexts}
As different contexts related to individuals' calendar events might have differing impacts while generating behavioral association rules for their scheduled events, we identify and determine the precedence of relevant contexts in our approach. For this purpose, we calculate the statistical property `information gain' that measures how well a particular context separates the training instances into targeted behavior classes, such as accept, reject, and missed phone call behavior for this particular problem. The context with the highest value of information gain is taken into account as the context of highest precedence. In order to determine the information gain value, we first need to calculate the entropy as well which is a measure of disorder or impurity. It reaches its maximum when the uncertainty is at a maximum and vice-versa. Formally entropy is defined as \cite{quinlan1993}:

$$H(DS)= -\sum_{x \in X} p(x)log_2p(x)$$

Where, $DS$ is the current data set for which entropy is being calculated, $X$ represents a set of classes in $DS$, $p(x)$ is the proportion of the number of elements in class $x$ to the number of elements in set $DS$. Based on entropy the formal definition of information gain as \cite{quinlan1993}:

$$IG(A, DS)= H(DS)-\sum_{t \in T} p(t)H(t)$$

Where, $H(DS)$ is the entropy of set $DS$, $T$ represents the subsets created from splitting set $DS$ by attribute $A$ such that $DS=\cup_{t \in T} t$, $p(t)$ is the proportion of the number of elements in $t$ to the number of elements in set $DS$, $H(t)$ is the entropy of subset $t$. 

Let's consider a sample calendar and phone log dataset of a mobile phone user X. For example, the contexts might be ranked as follows:

\noindent $Rank1: Event \; or \; Situation (S)\in \{meeting, lecture, seminar\}$ \\ 
$Rank2: Event \; Type (Tp) \in \{Recurring, Non$-$recurring\}$ \\
$Rank3: Relationship (R) \in \{boss, colleague, friend\}$ \\ 
Where, \\
$User \; behavior \in \{accept, reject, missed\}$ \\

\subsubsection{Designing Association Generation Tree}
An association generation tree is a typical tree structure that includes a root node, a number of branches, interior and/or leaf nodes. Each branch of the tree denotes a contextual test on a particular calendar event, and each node either interior or leaf represents the corresponding outcome containing the target phone call behavior class with confidence value of that calendar event based test. 

In order to build the association generation tree for various personalized calendar events, we follow a top-down tree structure, starting from a root node. The tree is partitioned into target behavior classes, such as accept, reject, and missed call behavior, distinguished by the values of the most relevant context related to the calendar events according to the precedence of contexts discussed above. Once the root node of the tree has been determined, the corresponding child nodes and its branches are generated. It then added to the tree with the associated contexts related to the calendar events and corresponding dominant behavior, e.g., reject, with determined confidence value, e.g., 85\%. As we aim to generate non-redundant association rules, we check whether a node is redundant or not, while creating a node in our calendar-based association generation tree. A child node in the resultant tree will be redundant, if same behavior class is also found in its parent node and satisfy individual's preferred confidence threshold as well.

\begin{figure}
	\centering
	\includegraphics[width=\linewidth, height = 5cm]{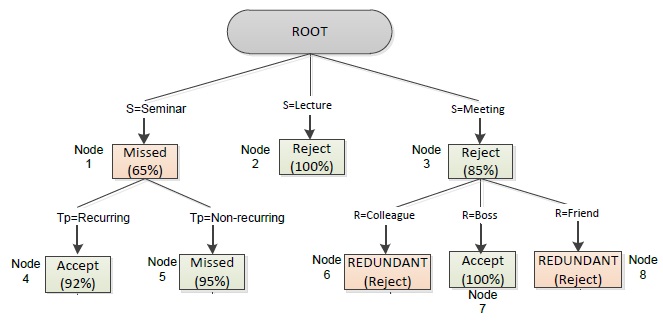}
	\caption{An example of an association generation tree (AGT) for various calendar events and corresponding behavior classes of a sample user.}
	\label{fig:tree}
\end{figure}

This association generation tree rule learning technique recursively adds new subtrees containing contextual test cases and corresponding behavioral classes as child nodes in order to grow the tree. We do not elaborate its children if a particular node has maximum value of confidence, i.e., 100\%. We dynamically continue this tree generating process according to the number of contexts in a given dataset. The final result of association generation tree is a multi-level tree with various nodes in which some of them are identified as `REDUNDANT' nodes. Figure \ref{fig:tree} shows an example of such an association generation tree containing `REDUNDANT' nodes by taking into account a number of contexts related to individual's calendar events and their phone call behaviors, when the minimum confidence preference is 80\%.

\subsubsection{Extracting Behavioral Association Rules}
Once the association generation tree has been designed, the behavioral association rules are then extracted by traversing the tree starting from the root node to the created interior or leaf nodes. To do this, we first identify the meaningful rule generating nodes from the tree. A node in the tree is taken into account as a meaningful rule generating node if it satisfies the confidence threshold preferred by an individual user and not labeled as `REDUNDANT'. The reason is that we take into account only non-redundant rules as the behavioral association rules in order to make the decision process effective. Thus, we are able to generate a concise set of behavioral association rules for individuals' calendar events using our tree-based approach, which are non-redundant. The following are examples of produced behavioral association rules from the tree, shown in Figure \ref{fig:tree}.

\noindent $R_1: {Lecture \Rightarrow Reject}$ (conf = 100\%, using node 2) \\
$R_2: {Meeting \Rightarrow Reject}$ (conf = 85\%, using node 3) \\
$R_3: {Seminar, Recurring \Rightarrow Accept}$ (conf = 92\%, using node 4) \\
$R_4: {Seminar, Non \; Recurring \Rightarrow Missed}$ (conf = 95\%, using node 5) \\
$R_5: {Meeting, Boss \Rightarrow Accept}$ (conf = 100\%, using node 7) \\

Rule $R_1$ states that the user always rejects the incoming calls (100\%) when she is in a lecture scheduled in her personal calendar, which is produced from node 2 in the designed association generation tree. Similarly, the other non-redundant behavioral association rules $R_2, R_3, R_4$ and $R_5$ for different calendar events are produced from the nodes 3, 4, 5, and 7 respectively according to the generated tree shown in Figure \ref{fig:tree}.

\section{Experiments}
\label{Experiments}
We have conducted a range of experiments to evaluate our personalized context-aware model, ``CalBehav'', for mining the behavioral association rules of individual mobile users for their various scheduled events. In this section, first we summarize the questions that we aim to answer by the experiments and describe the experimental setup. Then, we analyze the experimental results in different dimensions through answering these questions.

\subsection{Experimental Setup}
In these evaluation experiments, we aim to answer the following three questions:
\begin{itemize}
	\item Question 1: Are the produced event based rules personalized and reflect individual's behavioral patterns according to their preferences?
	\item Question 2: Does the data-driven method produce quality rules in terms of statistical measures?
	\item Question 3: How effective is our proposed machine learning based calendar behavioral model ``CalBehav'' relative to existing calendar-based approaches?
\end{itemize}

In answering these questions, we conduct experiments on real datasets of both calendar schedules and mobile phone data of different users. We have implemented our method in Java programming language and executed them on a Windows PC with an Intel Core I5 CPU (3.20GHz) and 8GB memory. In the following subsections, we discuss the experimental results that answer the above research questions.  

\subsection{Calendar and Mobile Phone Datasets}
We have conducted experiments on ten different datasets of different individuals collected by us. Each dataset contains both the phone call activities of an individual mobile phone user and her calendar information. To collect mobile phone data, we developed an Android mobile app which collects the user's real call log data on their mobile phones. Using our app, data was collected from different individual mobile users of different professions such as undergraduate students, post graduate students, university lecturers and industry professionals over the period of twelve months. We report the overall results of our experiments on these datasets and illustrate our approach with the detailed of experimental results of two individuals selected randomly from the above mentioned datasets.

\subsection{Benchmark Methods}
As we present data-driven approach in this paper, for the purpose of effectiveness comparison, we first select two baseline methods that utilize the calendar information to predict user's behavior for mobile communication. These are:
  
\begin{itemize}
	\item \textit{Existence of an event based approach}: To infer the behavior of cell phone users in mobile communication, researchers have studied about the existence of an event in individual's personal calendar. For instance, if an event exists in one's calendar, it is assumed that the user is unavailable to answer the incoming calls during that scheduled event \cite{khalil2005improving}. The details of such event existence based approaches are discussed in Section \ref{Related Work}. For comparison purposes, we denote this baseline method as BM1.
	
	\item \textit{Keyword based approach}: To infer the behavior of cell phone users in mobile communication, researchers have also studied about special keywords or the name of an event based approach. For instance, the system uses a collection of special keywords representing event names, in order to identify the behavioral variations for different type of events while generating rules \cite{dekel2009minimizing}. The details of such keyword or event name based approaches are discussed in Section \ref{Related Work}. For comparison purposes, we denote this baseline method as BM2.
\end{itemize}

\subsection{Evaluation Metrics}
To evaluate our data-driven approach for mining the behavioral rules of individuals for their various calendar events, we use the following metrics.

\begin{itemize}
	\item \textit{Accuracy and Coverage:} A rule R can be assessed by its coverage and accuracy \cite{han2011data}. Given a tuple, X, from a class labeled dataset, D, let $n_{covers}$ be the number of tuples covered by R;  $n_{correct}$ be the number of tuples correctly predicted by R; and $|D|$ be the number of tuples in D. We can define the accuracy and the coverage of R as -
	
	\begin{equation}
	Accuracy = \frac{n_{correct}}{n_{covers}} * 100\%
	\end{equation}	
	
	\begin{equation}
	Coverage = \frac{n_{covers}}{|D|} * 100\%
	\end{equation}	
	
	\item \textit{Rule Error Rate:} It measures the percentage of incorrect rules over the total number of rules that is determined by the best matching rules ($contexts \Rightarrow behavior$). If the number of incorrect rules is I and number of total rules is N then the formal definition of error rate is -
	
	\begin{equation}
	Error \; Rate = \frac{I}{N} * 100\%
	\end{equation}				
\end{itemize}

\subsection{Personalized Behavioral Rules}
To answer the above first question, in this section, we show sample individualized behavioral rules for various calendar events produced by our data-driven approach for two different users in Table \ref{Sample-mined-rules}. The behavioral rules are presented for User U1 and U2 respectively for various calendar events utilizing their individual's mobile phone data as an evidence, with a minimum confidence threshold of 75\%.

\begin{table*}
	\centering
	\caption{Sample behavioral rules for various scheduled events}
	\label{Sample-mined-rules}
	
	\begin{tabular}{|c|p{0.5cm}|p{10cm}|c|} \hline
		\bf Users & \bf Rule No & \bf \centering Behavioral Association Rules & \bf Confidence \\ \hline 
		& R1 & $EventName \rightarrow Meeting, EventType \rightarrow Nonrecurring
		\Rightarrow Behavior \rightarrow Reject$ & 100\% \\ \cline{2-4}	
		& R2 & $EventName \rightarrow Meeting, EventType \rightarrow Recurring
		\Rightarrow Behavior \rightarrow Reject$ & 82\% \\ \cline{2-4}
		U1 & R3 &$EventName \rightarrow Meeting, EventType \rightarrow Recurring, Relationship \rightarrow mother
		\Rightarrow Behavior \rightarrow Accept$ & 100\% \\ \cline{2-4}		
		& R4 & $EventName \rightarrow Practical, DayTime \rightarrow Wednesday[17:00-19:00]
		\Rightarrow Behavior  \rightarrow Accept$ & 75\% \\  \cline{2-4}	
		& R5 & $EventName \rightarrow Practical, DayTime \rightarrow Monday[12:00-16:00]
		\Rightarrow Behavior  \rightarrow Reject$ & 100\% \\  \hline 	
		
		& R6 & $EventType \rightarrow Nonrecurring
		\Rightarrow Behavior \rightarrow Accept$ & 75\% \\ \cline{2-4}
		& R7 & $EventType \rightarrow Nonrecurring, Relationship \rightarrow unknown 
		\Rightarrow Behavior \rightarrow Missed$ & 100\% \\ \cline{2-4}		
		U2 & R8 & $EventName \rightarrow Class
		\Rightarrow Behavior \rightarrow Reject$ & 100\% \\ \cline{2-4}
		& R9 & $EventName \rightarrow Busy, DayTime \rightarrow Friday[11:00-15:00]
		\Rightarrow Behavior  \rightarrow Accept$ & 100\% \\  \cline{2-4}
		& R10 & $EventName \rightarrow Busy, DayTime \rightarrow Sunday[18:00-20:00]
		\Rightarrow Behavior  \rightarrow Reject$ & 100\% \\  \hline
	\end{tabular}
\end{table*}

If we observe the rules produced in Table \ref{Sample-mined-rules}, we see that people are clearly well differ from each other for some events, in how they respond to incoming calls during their scheduled events. For example, rule R1 states that user U1 always (100\%) rejects the incoming calls during the nonrecurring event meeting while rule R6 states that user U2 accepts most of the incoming calls (75\%) during the similar event type nonrecurring. Such individualized behavior-oriented rules produced by our data-driven calendar-based model provide an evidence of personalization in mining their behavioral association rules for various calendar events according to their unique behavioral patterns. In addition, a particular individual user may respond differently to what category of event is scheduled in the calendar. For example, rule R4 states that the user U1 accepts most of the incoming calls (75\%) during another event practical that represents lab work, which is different from the event meeting. Even a particular individual may respond differently for like or similar events. For example, the behavior of user U1 in rule R4 and the behavior in rule R5 are clearly different for like event practical. Besides these event related contexts, social relational context has a great impact on their personalized behavioral association rules for calendar events. For example, Rule R3 states that, the user U1 always (100\%) accepts the incoming phone calls during the recurring event meeting, if the calls come from her mother (social relationship). However, this behavior is different for the similar event, which is stated in Rule R2. As the behavior of different individuals are not identical in the real world, our data-driven calendar based model CalBehav captures the actual dominant behavior for various calendar events of individuals utilizing their smartphone log data.

\subsection{Effect on Coverage and Accuracy}
\begin{figure}[htbp!]
	\centering
	\includegraphics[width=\linewidth, keepaspectratio]{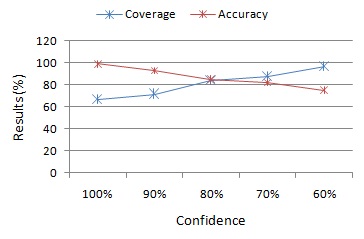}
	\caption{Effect of confidence on coverage and accuracy of an individual's smartphone dataset DS-U1.}
	\label{fig:Dataset-1}
\end{figure}

\begin{figure}[htbp!]
	\centering
	\includegraphics[width=\linewidth, keepaspectratio]{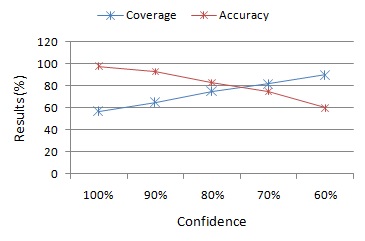}
	\caption{Effect of confidence on coverage and accuracy of an individual's smartphone dataset DS-U2.}
	\label{fig:Dataset-2}
\end{figure}

To answer the second question, in this experiment, we show the effect of confidence on coverage and accuracy of the discovered personalized behavioral rules produced by our calendar-based model CalBehav. For this, we first illustrate the detailed outcomes, in terms of incoming call responses for various calendar events, by varying the conference threshold from 100\% (maximum) below to 60\% (lowest) for different individuals. Since by the definition mentioned in Section \ref{Problem-Statement}, confidence is associated to rules' strength or accuracy, we are not interested to take into account below 60\% value as confidence preference. To show the effect of confidence on coverage and accuracy, Figure \ref{fig:Dataset-1} and Figure \ref{fig:Dataset-2} show the relationship between coverage and accuracy for different confidence threshold utilizing individual's datasets DS-U1 and DS-U2 respectively.

If we observe Figure \ref{fig:Dataset-1} and Figure \ref{fig:Dataset-2} , we can see that for 100\% confidence threshold, the resultant accuracy of the discovered rules is high but the corresponding coverage is low. The reason is that using 100\% confidence threshold, our calendar-based model generates a set of rules with like behavior for a particular event, which represent consistent behavior of an individual without any variation in her behavior for that event. However, such consistency in behavior modeling may not found for many events because of one's day-to-day situations in the real world life. Thus the coverage value is low as it extracts fewer rules for this given maximum threshold of confidence preference. On the other hand, if the confidence preference becomes low, the accuracy decreases but the coverage increases. Because it is then satisfied by more rules and output all the rules above this low threshold. Thus, the total support in rules increases and coverage increases as well. The setting of this confidence threshold for creating rules will vary according to an individual's preference.

Typically, higher accuracy results in a lower coverage and vice-versa. Using a higher confidence threshold results in higher accuracy but lower coverage, and using a lower confidence threshold results in lower accuracy but higher coverage. Further, we allow users to configure the coverage-accuracy trade off based on their individual preferences (e.g., say 80\% confidence). 

\subsection{Effectiveness Comparison}
To answer the third question, first we calculate the error rate of the personalized behavioral rules produced for individual's scheduled events by our calendar-based approach Calbehav. For this purpose, we utilize a 5-fold cross validation technique to evaluate each context-aware test cases. In $k$ fold cross-validation, the initial data is randomly partitioned into $k$ mutually exclusive subsets or ``folds'', $d_1,d_2,...,d_k$, each of which has an approximately equal size of data. Training and testing are performed $k$ times in order to output an average result of all the iterations. In iteration $i$, the partition $d_i$ is reserved as the test set, and the remaining partitions are collectively used to train the model. To be specific, we first randomly divide individual's phone log data into five equal parts, then we use each part as the test data while using the other four parts as the training data in five test rounds. Finally, we calculate the error rate using the above definition and report the average error rate of the five runs. This average value is then used as the error rate of our approach for each individual user.

\begin{figure}[h]
	\centering
	\includegraphics[width=\linewidth, keepaspectratio]{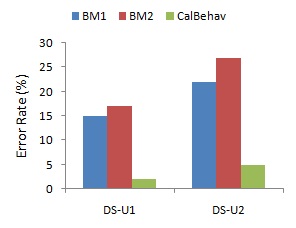}
	\caption{Effectiveness comparison in terms of error rates (\%) of different calendar-based user behavior models utilizing individual's datasets.}
	\label{fig:error-rate-individuals}
\end{figure}

\begin{figure}[h]
	\centering
	\includegraphics[width=\linewidth, keepaspectratio]{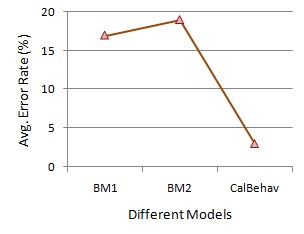}
	\caption{Effectiveness comparison in terms of average error rates (\%) of different calendar-based user behavior models utilizing a collection of datasets.}
	\label{fig:error-rate-avg}
\end{figure}

For the purpose of comparing the effectiveness of our calendar-based approach, Figure \ref{fig:error-rate-individuals} shows the relative comparison of the error rate for the produced behavioral rules by different approaches for two different datasets. In addition to the individual's comparison, Figure \ref{fig:error-rate-avg} shows the relative comparison of the average error rate of all the datasets. As we have flexibility in confidence, the results are shown using the confidence setting 80\% by taking into account the trade-of between coverage and accuracy shown in the above. In general, error rate increases with the decrease of confidence threshold and vice-versa.

From Figure \ref{fig:error-rate-individuals} and \ref{fig:error-rate-avg}, we find that our data-driven calendar-based approach CalBehav consistently outperforms previous calendar-based approaches. The main reason is that our CalBehav model can identify the actual dominant behavior for each individual user more accurately based on their past call activity records in phone log. Thus, the corresponding produced set of behavioral association rules for their calendar events become more meaningful because of their low error rate in predicting context-aware test cases, shown above. On the other hand, in the existing calendar-based approaches, the behavioral rules are static and non-personalized, they cannot represent accurately user's diverse behavior in various day-to-day situations for various events scheduled in their personal calendar. In contrast, our data-driven calendar based model CalBehav resolves these issues while modeling individual's behavior and better captures their behavioral patterns for incoming mobile communications for their various scheduled events.

\section{Discussion}
\label{Discussion}
Overall, our machine learning based personalized calendar behavioral model CalBehav is fully personalized and can assist the application developers for building an intelligent mobile interruption management system for the end users. Compared to the existing calendar based approaches, the error rate of the discovered rules is decreased when our approach is used, as shown in Figure \ref{fig:error-rate-individuals} and Figure \ref{fig:error-rate-avg}. The following are some key discoveries from our calendar-based study.

In the real world, the behavioral rules of individual mobile phone users for their various calendar events are not static. In our experiments, we have discovered different behavior for different scheduled events of individuals, according to their past call response records in their mobile phone log. The reason is that different users have different preferences and therefore different behavioral patterns and rules as well. 
	
Interestingly, our study showed that only calendar entries are not enough to assume one's actual behavior in mobile communication. Thus, incorporating another data source time-series `smartphone log' as a behavioral evidence plays an important role in capturing the behavioral patterns of individuals. In Figure \ref{fig:behavioral-evidence}, we have shown users behave differently for different scheduled events in their real world life. Thus, a static event-behavior mapping for individuals becomes meaningless and not appropriate to assume personalized behavior for their personal calendar events. 
	
Another important finding of our study is that the data-driven approach CalBehav produces quality rules in terms of statistical measures, i.e., coverage and accuracy. Moreover, it produces different sets of rules for different confidence thresholds preferred by individuals, which may vary from user-to-user in the real world. Thus, the rules produced by our approach is highly applicable in the real word applications because of its effectiveness to capture the actual dominant behavior of individuals in various context-aware test cases with low error rate. 
	
In this work, we have concentrated on producing meaningful behavioral association rules for calendar events of individual users for the purpose of managing incoming mobile communications in their different day-to-day situations. However, our data-driven rule mining approach based on association generation tree can also be applicable to other application domains like intelligent cyber security systems with relevant multi-dimensional contexts, particularly, rule-based context-aware access control systems \cite{kayes2012icaf} \cite{kayes2018policy}. While our approach is able to capture the behavior patterns utilizing mobile phone log data and produce meaningful personalized rules, our current approach is not applicable at cold start, where no data exists in phone log or a new phone. Furthermore, to verify the usefulness of the mined rules in action, it would be necessary to incorporate these rules with an interruption management system and carry out a qualitative analysis, including a usability study for the rules and corresponding rule-based system.

\section{Conclusion and Future Work}
\label{Conclusion and Future Work}
In this paper, we have presented a data-driven calendar based personalized behavioral model CalBehav that generates a set of concise behavioral rules of individuals for their various calendar events. Our CalBehav model is also known as evidence-based user behavior model as we have used individuals' past call handing records to infer their actual behavior during the scheduled events. In order to achieve our goal, we have designed a dynamic event-behavior mapping based on relevant contextual information utilizing both the calendar and phone log datasets of individuals. Consequently, behavioral call response rules are formulated using rule based machine learning technique for each scheduled event in the calendar, which can be used to assist the end users in an intelligent mobile interruption management system. The rules are generated based on the contiguous dominant behavior of the users according to their own preferences and will be different from event-to-event or user-to-user to truly reflect their actual behavior. Experiments on real datasets collected from multiple sources show that our calendar-based context-aware model is more effective compared to existing calendar-based approaches to infer individual's behavioral rules for their various calendar events.

To develop a data-driven intelligent system for mobile phone users based on the presented CalBehav model and to assess their satisfaction with the generated behavioral rules for calendar events by conducting a user survey could be a future work.

\section*{Acknowledgment}
\label{Acknowledgment}

The authors would like to thank the users who provided their datasets and the administrative staff of Swinburne University of Technology, Melbourne, Australia, for their support while doing this work and experiment in their post-graduate research lab.

\bibliographystyle{compj}
\bibliography{CalendarBehav-bibfile}

\end{document}